\documentclass[aps,twocolumn,a4paper]{revtex4}
\usepackage{amsmath}
\begin{document}
\title{Relating monomer to center-of-mass distribution functions in
macromolecular fluids: extension to arbitrary systems}

\author{V. Krakoviack}

\affiliation{Laboratoire de Chimie, UMR CNRS 5182, \'Ecole Normale
Sup\'erieure de Lyon, 46 All\'ee d'Italie, 69364 Lyon Cedex 07,
France}

\begin{abstract}
In the framework of the polymer reference interaction site model
(PRISM) extended to incorporate the centers of mass (CM) of the groups
of monomers constitutive of the macromolecules as auxiliary sites, a
general relationship between the monomer-monomer and CM-CM pair
correlation functions is derived for arbitrary mixtures of
macromolecules of arbitrary structure. It extends previous results for
homopolymer fluids and should be useful to develop semi-analytic
coarse-grained descriptions of complex fluids.
\end{abstract}

\maketitle

Following the recent successes of a mesoscale coarse-graining strategy
for polymeric systems, in which the microscopic chain configurations
are mapped onto those of a fluid of mesoscopic particles by the
introduction of effective state-dependent pair interactions between
the centers of mass (CM) of the polymer coils derived from the CM-CM
pair distribution function of the original system
\cite{bollouhanmei01jcp}, a new interest has emerged for the
development of theoretical approaches able to predict these CM-CM
correlations from first principles in a wide range of thermodynamic
conditions.  Recently, a very simple route towards such theories has
been opened with the derivation, in the framework of the polymer
reference interaction site model (PRISM) \cite{schweizer1997}, of an
approximate relation, involving only single coil structural
quantities, between the monomer-monomer and CM-CM pair correlation
functions of a one-component homopolymer fluid
\cite{krahanlou02el}. Indeed, with this formula, the predictions of
which compare favorably with simulation data in wide density and
solvent quality ranges
\cite{krahanlou02el,krarothan04jpb,aicchobasfuc04pre}, any theory for
monomer-monomer correlations can virtually be transformed into a
theory for CM-CM correlations. This has been put to good use to derive
effective CM-CM pair interactions starting from a very simple integral
equation theory, with results in good agreement with simulation data,
at least in the good solvent regime \cite{krarothan04jpb}.

Lately, Guenza and coworkers have extended the relation derived in
Ref.~\cite{krahanlou02el} to binary polymer mixtures and, following a
procedure analogous to that of Ref.~\cite{krarothan04jpb}, they have
computed CM-CM pair distribution functions for melts and binary
blends, with results in fair agreement with molecular dynamics data
\cite{yatsamnemgue04prl}, thus demonstrating the potential
applicability of the theory proposed in Ref.~\cite{krahanlou02el} to
more complex macromolecular systems. In this Note, we seize the
opportunity provided by this work and by the recent extension of the
above coarse-graining strategy to diblock copolymer solutions
\cite{addhankralou05mp}, to present the generalization of the results
of Refs.~\cite{krahanlou02el} and \cite{yatsamnemgue04prl} to
arbitrary mixtures of macromolecules of arbitrary structure.

As in Refs.~\cite{krahanlou02el} and \cite{yatsamnemgue04prl}, the
sought-for relation results from the application of the PRISM theory,
now in its general multicomponent formulation, to macromolecular fluid
systems in which two categories of sites are formally made to coexist:
on the one hand, ``real'' or physical sites, which actually interact
and whose interactions determine the microscopic structure of the
fluid, and, on the other hand, auxiliary non-interacting sites which
are only introduced to tag specific points of interest, typically
following from a simple geometric definition like the CMs.
 
In this approach, a macromolecular species $x$, present in the mixture
at number density $\rho_x$, is assumed to consist of $n_x$ classes of
rigorously or approximately equivalent physical sites denoted by
$\mathbf{x}\equiv\{x_1,\ldots,x_{n_x}\}$ and $m_x$ classes of
rigorously or approximately equivalent auxiliary sites denoted by
$\mathbf{X}\equiv\{X_1\equiv x_{n_x+1}, \ldots,X_{m_x}\equiv
x_{n_x+m_x}\}$. We denote by $N_{x_i}$, $1\leq i\leq n_x+m_x$, the
number of sites of class $x_i$ per macromolecule.

Then, in reciprocal space at wave vector $\mathbf{q}$, the PRISM
equations read, in matrix form \cite{schweizer1997},
\begin{equation}
\mathbf{H}(\mathbf{q})= \mathbf{W}(\mathbf{q}) \mathbf{C}(\mathbf{q})
[\mathbf{W}(\mathbf{q})+\mathbf{H}(\mathbf{q})].
\end{equation}
$\mathbf{W}(\mathbf{q})$ is the symmetric matrix of intramolecular
structure factors, or form factors, defined as
\begin{equation}
W_{x_i x_j}(\mathbf{q})= \sum_{k=1}^{N_{x_i}} \sum_{l=1}^{N_{x_j}}
\langle e^{i\mathbf{q}\cdot(\mathbf{r}_k^{x_i}-
\mathbf{r}_l^{x_j})}\rangle,
\end{equation}
where $x_i$ and $x_j$, $1\leq i,j\leq n_x+m_x$, are classes of real or
auxiliary sites belonging to the molecules of species $x$ and
$\mathbf{r}_k^{x_i}$ and $\mathbf{r}_l^{x_j}$ are the positions of the
sites of these classes on one macromolecule. $\mathbf{W}(\mathbf{q})$
is block diagonal and, separating the coefficients according to the
real or auxiliary nature of the involved sites, the diagonal blocks
can be written $\left(
\begin{smallmatrix} \mathbf{W}_{\mathbf{xx}}(\mathbf{q}) &
\mathbf{W}_{\mathbf{xX}}(\mathbf{q}) \\
\mathbf{W}_{\mathbf{Xx}}(\mathbf{q}) &
\mathbf{W}_{\mathbf{XX}}(\mathbf{q}) \end{smallmatrix} \right)$.
$\mathbf{H}(\mathbf{q})$ and $\mathbf{C}(\mathbf{q})$ are
non-symmetric matrices simply related to the site-averaged total and
direct correlation functions $h_{x_i y_j}(\mathbf{q})$ and $c_{x_i
y_j}(\mathbf{q})$ through the relations
\begin{equation} \label{relation}
H_{x_i y_j}(\mathbf{q})=\rho_x N_{x_i} N_{y_j} h_{x_i
y_j}(\mathbf{q}), \ C_{x_i y_j}(\mathbf{q})=\rho_x c_{x_i
y_j}(\mathbf{q}),
\end{equation}
and, using the same separation of site classes as above, they are
found to consist of blocks of the form $\left( \begin{smallmatrix}
\mathbf{H}_{\mathbf{xy}}(\mathbf{q}) &
\mathbf{H}_{\mathbf{xY}}(\mathbf{q}) \\
\mathbf{H}_{\mathbf{Xy}}(\mathbf{q}) &
\mathbf{H}_{\mathbf{XY}}(\mathbf{q}) \end{smallmatrix} \right)$ and
$\left( \begin{smallmatrix} \mathbf{C}_{\mathbf{xy}}(\mathbf{q}) &
\mathbf{C}_{\mathbf{xY}}(\mathbf{q}) \\
\mathbf{C}_{\mathbf{Xy}}(\mathbf{q}) &
\mathbf{C}_{\mathbf{XY}}(\mathbf{q}) \end{smallmatrix} \right)$,
respectively.

We now partially close the set of PRISM equations by setting all
direct correlation functions involving auxiliary sites identically to
zero. This approximation is motivated on physical grounds by the fact
that these sites are strictly non interacting and that all
correlations in which they are involved can only originate from
intramolecular correlations and direct interactions between physical
sites. This results in a specific structure of the direct correlation
matrix, consisting of simple blocks $\left(
\begin{smallmatrix} \mathbf{C}_{\mathbf{xy}}(\mathbf{q}) & \mathbf{0}
\\ \mathbf{0} & \mathbf{0} \end{smallmatrix} \right)$.

It is now a matter of simple matrix algebra to show that 
\begin{gather} \label{first}
\mathbf{H}_{\mathbf{Xy}}(\mathbf{q}) =
\mathbf{W}_{\mathbf{Xx}}(\mathbf{q})
\mathbf{W}^{-1}_{\mathbf{xx}}(\mathbf{q})
\mathbf{H}_{\mathbf{xy}}(\mathbf{q}), \\
\mathbf{H}_{\mathbf{XY}}(\mathbf{q}) =
\mathbf{W}_{\mathbf{Xx}}(\mathbf{q})
\mathbf{W}^{-1}_{\mathbf{xx}}(\mathbf{q})
\mathbf{H}_{\mathbf{xY}}(\mathbf{q}),
\end{gather}
where $\mathbf{W}^{-1}_{\mathbf{xx}}(\mathbf{q})$ denotes the inverse
matrix of $\mathbf{W}_{\mathbf{xx}}(\mathbf{q})$, and, by
transposition and multiplication by $\rho_y/\rho_x$,
\begin{gather}
\mathbf{H}_{\mathbf{yX}}(\mathbf{q}) =
\mathbf{H}_{\mathbf{yx}}(\mathbf{q})
\mathbf{W}^{-1}_{\mathbf{xx}}(\mathbf{q})
\mathbf{W}_{\mathbf{xX}}(\mathbf{q}), \\
\mathbf{H}_{\mathbf{YX}}(\mathbf{q}) =
\mathbf{H}_{\mathbf{Yx}}(\mathbf{q})
\mathbf{W}^{-1}_{\mathbf{xx}}(\mathbf{q})
\mathbf{W}_{\mathbf{xX}}(\mathbf{q}).
\end{gather} 
Combining these equations, it results that
\begin{equation} \label{last}
\mathbf{H}_{\mathbf{XY}}(\mathbf{q}) =
\mathbf{W}_{\mathbf{Xx}}(\mathbf{q})
\mathbf{W}^{-1}_{\mathbf{xx}}(\mathbf{q})
\mathbf{H}_{\mathbf{xy}}(\mathbf{q})
\mathbf{W}^{-1}_{\mathbf{yy}}(\mathbf{q})
\mathbf{W}_{\mathbf{yY}}(\mathbf{q}).
\end{equation}
Note that there is an obvious simplification by $\rho_x$ on both sides
of this equation, such that no explicit dependence in the densities is
present.

If the auxiliary sites are the CMs of monomer groups constitutive of
the macromolecules, then Eq.~\eqref{last} is the generalized form of
the expressions derived in Refs.~\cite{krahanlou02el} and
\cite{yatsamnemgue04prl}. It is flexible enough to accommodate various
ways of building coarse-grained representations of the macromolecules
based on the CMs of their monomers.  Indeed, for physical reasons or
analytical convenience, one might choose to introduce the CMs of the
sites of each class, leading to a coarse-grained representation with
$m_x=n_x$ sites for species $x$, as it was recently done for diblock
copolymers ($n_x=2$) \cite{addhankralou05mp}, or one might find useful
to work with the global CM of the macromolecules only, hence $m_x=1$,
or any intermediate grouping of non-equivalent site classes, then $m_x
< n_x$. Of course, different choices for the various components of the
mixture can be mixed. In the following, we explicitly discuss some of
the simplest examples.

The first one, relevant for homopolymer or colloid-homopolymer
mixtures, corresponds to $n_x=m_x=1$ for all species. Then all
matrices in Eq.~\eqref{last} reduce to scalars and one finds
\begin{equation}
h_{X_1Y_1}(\mathbf{q}) =
\frac{W_{X_1x_1}(\mathbf{q})W_{Y_1y_1}(\mathbf{q})}
{W_{x_1x_1}(\mathbf{q})W_{y_1y_1}(\mathbf{q})} N_{x_1} N_{y_1}
h_{x_1y_1}(\mathbf{q}),
\end{equation}
as shown in Refs.~\cite{krahanlou02el} and \cite{yatsamnemgue04prl},
where a slightly different normalization was used for the form
factors.

More generally, for any species $x$ comprising only one type of
monomers ($n_x=1$) in an arbitrary mixture, any pair correlation
function $h_{x_1y}(\mathbf{q})$ involving these monomers can be
transformed into the corresponding pair correlation function
$h_{X_1y}(\mathbf{q})$, involving the CMs of these monomers, using the
simple relation
\begin{equation}
h_{X_1y}(\mathbf{q}) = \frac{W_{X_1x_1}(\mathbf{q})}
{W_{x_1x_1}(\mathbf{q})} N_{x_1} h_{x_1y}(\mathbf{q}),
\end{equation}
where $y$ denotes any type of site, real or auxiliary, on any
macromolecular species, including sites of species $x$ itself.

We now consider the case of a one-component fluid with $n_x>1$ and
$m_x=1$, i.e., the macromolecules are made of different types of
monomers, but a coarse-grained description with their global CMs as
the reference point is sought. Then, expanding Eq.~\eqref{last},
one finds
\begin{widetext}
\begin{equation}
h_{X_1X_1}(\mathbf{q}) = \sum_{i,j,k,l=1}^{n_x} W_{X_1x_i}(\mathbf{q})
[\mathbf{W}^{-1}_{\mathbf{xx}}]_{x_ix_j}(\mathbf{q}) N_{x_j} 
h_{x_j x_k}(\mathbf{q}) N_{x_k}
[\mathbf{W}^{-1}_{\mathbf{xx}}]_{x_kx_l}(\mathbf{q})
W_{x_lX_1}(\mathbf{q}),
\end{equation}
\end{widetext}
a result which might be useful to study copolymer or star polymer
fluids.

In conclusion, we have derived fully general approximate relations
between the monomer and CM pair correlation functions in
macromolecular fluid systems. They extend previous results obtained
for simple homopolymer fluids which have been found to compare very
favorably with computer simulation data. With these equations,
following the lines of the work done in Ref.~\cite{krarothan04jpb}, a
systematic investigation of effective pair interactions in
macromolecular fluids seems achievable, at a modest computational cost
and with at least semi-quantitative accuracy. Work along this line is
in progress.

\end{document}